\documentclass[%
 reprint,
groupedaddress,
 amsmath,amssymb,
 superscriptaddress,
 aps,
 prl,
 longbibliography
]{revtex4-2}

\usepackage{graphicx}
\usepackage{dcolumn}
\usepackage{bm}
\usepackage{braket}
\usepackage{physics}
\usepackage{ulem}

\usepackage[unicode=true]{hyperref}
\usepackage{xcolor}

\hypersetup{
  colorlinks = true, 
  linkcolor = blue, 
  filecolor = blue, 
  urlcolor = blue, 
  citecolor = blue 
}

\usepackage{graphicx}
\usepackage{dcolumn}
\usepackage{bm}

\newcommand{\magenta}[1]{\textcolor{magenta}{#1}}

\begin{document}

\title{Strain-induced quantum oscillation in Kitaev spin liquid with Majorana-Fermi surface}

\author{Takayuki Yokoyama}
\affiliation{Quantum Matter Program, Graduate School of Advanced Science and Engineering, Hiroshima University,
Higashihiroshima, Hiroshima 739-8530, Japan}

\author{Yasuhiro Tada}
\email[]{ytada@hiroshima-u.ac.jp}

\affiliation{Quantum Matter Program, Graduate School of Advanced Science and Engineering, Hiroshima University,
Higashihiroshima, Hiroshima 739-8530, Japan}
\affiliation{Institute for Solid State Physics, University of Tokyo, Kashiwa 277-8581, Japan}

\begin{abstract}
We theoretically study Landau quantization of itinerant Majorana quasiparticles induced by lattice strain in a Kitaev spin liquid with Majorana-Fermi surfaces. We consider the isotropic spin-1/2 Kitaev model on the honeycomb lattice with a perturbation such as a staggered Zeeman field and an electromagnetic field, which generates small Majorana-Fermi surfaces near the Dirac points. By introducing triaxial strain, we create an effective vector potential that couples to Majorana fermions and leads to Landau quantization. Our calculations show that the low-energy spectrum forms discrete pseudo–Landau levels of the Majorana-Fermi surface. We further demonstrate that the strain-induced effective vector potential gives rise to pronounced quantum oscillations of the density of states and the specific heats at very low temperatures in close analogy to the de Haas-van Alphen effect for charged electrons in metals. These results indicate that Landau-quantization–driven “Majorana quantum oscillations” can serve as a probe of the charge-neutral Majorana-Fermi surface in Kitaev materials.
\end{abstract}

\maketitle

\noindent\magenta{\it Introduction.}
Quantum spin liquids are exotic quantum phases that do not exhibit long-range magnetic order even at the lowest temperatures because of strong quantum fluctuations, and instead host fractionalized spin excitations \cite{Savary_2017}. These properties have made them a central topic in condensed-matter physics in recent years. The paradigmatic example is the $S=1/2$ Kitaev model on a honeycomb lattice \cite{KITAEV20062}. This model is one of exactly solvable spin liquid models that has been extensively studied. In the Majorana representation, the spin degrees of freedom fractionalize into itinerant Majorana fermions with the Dirac-type linear dispersion and gapped $\mathbb{Z}_2$ fluxes (visons). To understand these unconventional excitations, intensive efforts have been made on both experimental and theoretical fronts. On the experimental side, candidate materials include in particular $\alpha$-RuCl$_3$, where a half-integer quantized thermal Hall effect observed in a high Zeeman magnetic field has been interpreted as strong evidence for Majorana fermions characteristic of a Kitaev spin liquid \cite{matsuda2025, Yokoi2021, Bruin2022}. On the theoretical side, many studies have explored finite-temperature properties, perturbations to the Kitaev model such as Heisenberg and $\Gamma$ interactions, and the effects of impurities and other perturbations
\cite{TREBST20221,PRL.113.197205, PRB.92.115122, PRL.119.127204, PRL.105.027204, PRL.112.077204, PRL.117.037202, PRX.11.011034}.

Recently, phases with Majorana-Fermi surfaces have attracted particular attention. The Majorana-Fermi surface was first studied in the three dimensional Kitaev models and examined extensively also for two dimensions~\cite{PRB.89.235102, PRL.115.177205}.
For example, in the Kitaev model under a uniform Zeeman magnetic field, an intermediate spin liquid phase with Majorana-Fermi surfaces has been theoretically reported between the chiral spin liquid phase and the spin-polarized phase \cite{PRB.97.241110,PRB.98.014418,jiang2018_arxive, TREBST2019,pnas.1821406116, Zhu2025, PRB.111.L100402}.
It has been also pointed out that a staggered Zeeman magnetic field in the Kitaev model generates small Majorana-Fermi surfaces in the vicinity of the Dirac points \cite{PRB.99.224409}. This scenario was originally discussed for possible coexistence of the zigzag antiferromagnetic order and the spin liquid in $\alpha$-RuCl$_3$, but may be relevant also to the Kitaev spin liquid candidates with weak antiferromagnetism such as Cu$_2$IrO$_3$~\cite{PRL.102.017205, PRB.100.094418, doi:10.1021/jacs.7b06911, PRX.9.031047} and A$_3$LiIr$_2$O$_6$ (A $=$ Cu, Ag)~\cite{doi.org/10.1002/pssb.202100146, PRL.108.127203, PRL.123.237203, C9DT01789E},
and a hetrostructure of a Kitaev material and an antiferromagnet~\cite{PRB.105.165152}.
The Majorana-Fermi surface can also emerge under a static electromagnetic field due to the cross term of the magnetic and electric fields~\cite{PRB.103.134444}.
Experimentally, scanning tunneling microscopy (STM) measurements on $\alpha$-RuCl$_3$ have revealed Friedel-oscillations that can be associated with Fermi surfaces, and have been discussed as evidence for locally induced Majorana-Fermi surfaces created by the STM current \cite{PRX.14.041026}.

In this work, we propose quantum oscillations induced by a strain-generated pseudo-magnetic field as a new route to probe the properties of Majorana-Fermi surfaces. Because charge-neutral Majorana quasiparticles do not couple to conventional Zeeman magnetic fields, they do not show ordinary Landau quantization. However, when lattice strain is introduced and an effective vector potential is generated that couples to the Majorana fermions near the Dirac points, it is known that pseudo–Landau quantization can occur even in the Kitaev model \cite{PRL.116.167201, JPSJ.92.114705, PRB.105.085147}. Here we extend this idea to the case where Majorana-Fermi surfaces appear near the Dirac points in the Kitaev model with a perturbation, and study the effects of triaxial strain. Our numerical calculations show that discrete pseudo–Landau levels are formed in the presence of the Majorana-Fermi surfaces state under the strain. By analyzing the density of states and the specific heat, we demonstrate “Majorana quantum oscillations" analogous to the de Haas–van Alphen effect in metals under magnetic fields. Our results point to a new experimental probe that can directly detect Majorana-Fermi surfaces.

\begin{figure}[t]
  \centering
  \includegraphics[width=1.0\columnwidth]{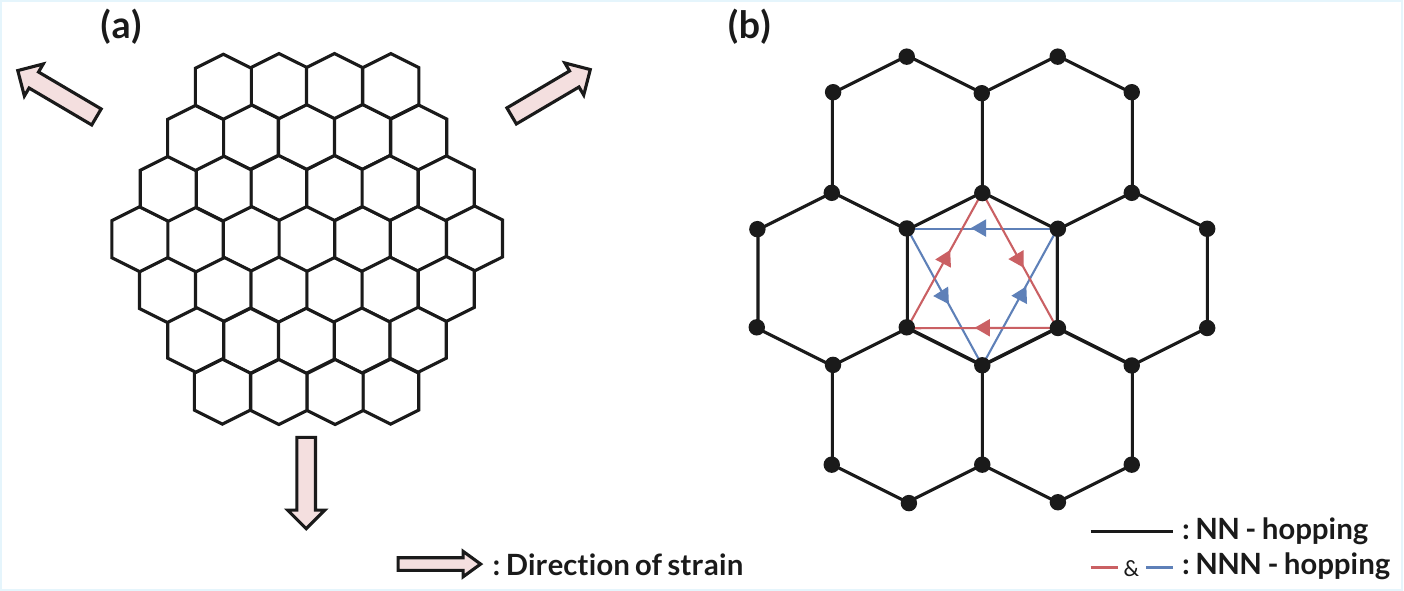}
  \caption{(a) Geometry of the hexagonal flake with zigzag boundaries used in the calculations ($r=4$). 
The arrows indicate the directions of the applied triaxial strain. 
(b) Kitaev model in a staggered Zeeman magnetic field in the Majorana representation. 
Black lines denote nearest-neighbor hoppings. 
Red (blue) bonds denote next-nearest-neighbor hoppings with $d_{ij}=+1$ ($d_{ij}=-1$), respectively [see Eq.~\eqref{eq::majorana_Hamiltonian}].}
\label{fig::majorana_scketch}
\end{figure}

\begin{figure}[t]
  \centering
  \includegraphics[width=1.0\columnwidth]{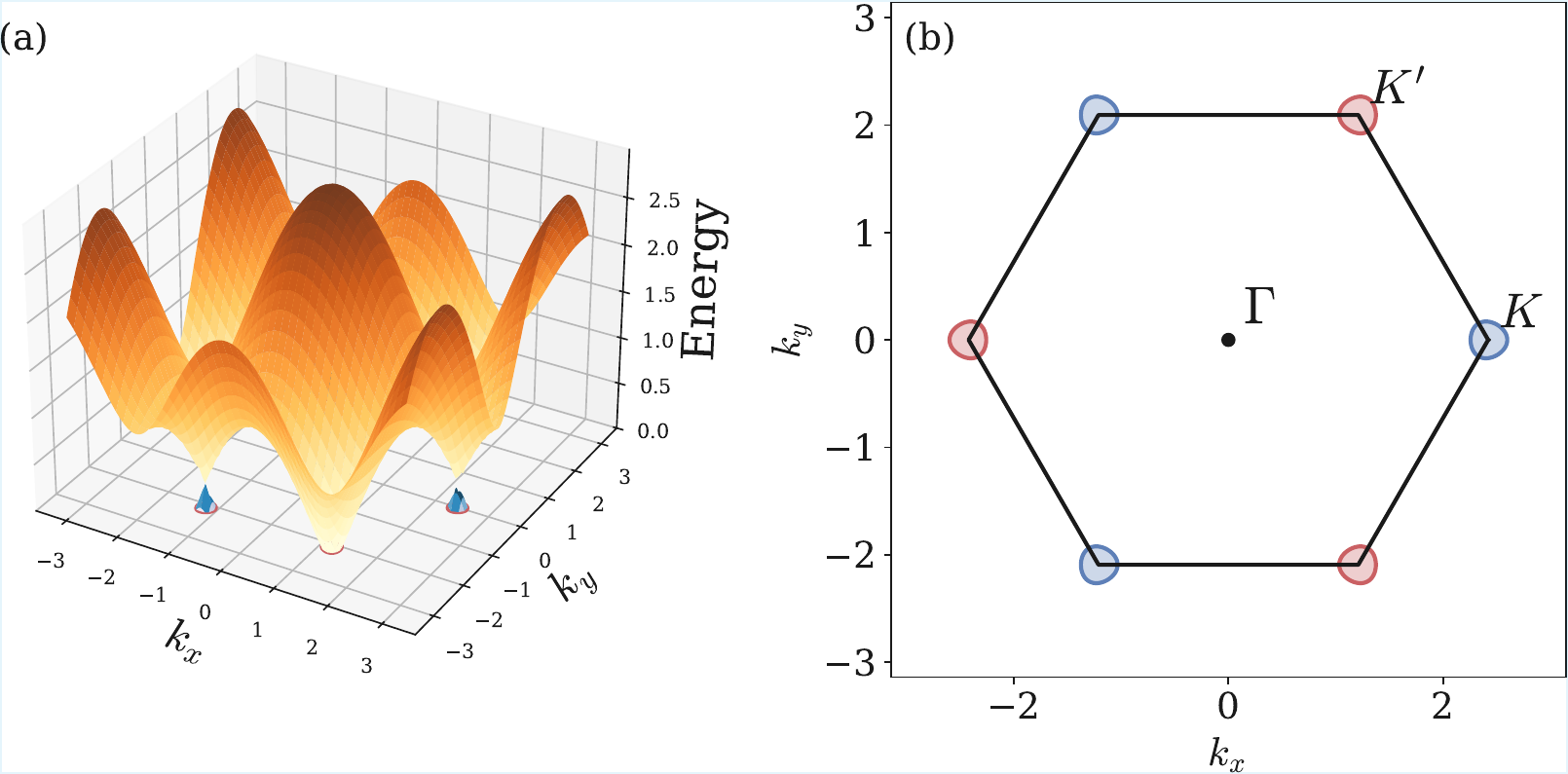}
  \caption{Energy spectrum at $h_{\mathrm{st}}^3=0.05$ in momentum space.
(a) The dispersion as a function of the momenta.
(b) The Majorana-Fermi surface. The blue region corresponds
to a hole-like pocket of Majorana fermions, while the red region does to an electron-like pocket.}
  \label{fig:BZ}
\end{figure}

\noindent\magenta{\it Kitaev model with a perturbation.}
We consider the isotropic $S=1/2$ Kitaev model,
\begin{align}
    \mathcal{H}_K &=  \sum_{\alpha} \sum_{\ev{i,j}_\alpha} J_{ij} S^{\alpha}_i S^{\alpha}_j ,
    \label{eq::Hamiltonian}
\end{align}
where $S^{\alpha}_i$ ($\alpha = x, y, z$) denotes the $S=1/2$ spin operator at site $i$.
As mentioned in the introduction, the honeycomb lattice Kitaev model can have a spin liquid ground state with a Majorana-Fermi surface in the presence of perturbations $\mathcal{H}_{\mathrm{pert}}$. 
For example, the electric ($E$) and magnetic ($B$) fields can induce an effective perturbation term at low energies $\mathcal{H}_{\mathrm{pert}}\sim\sum_{ijk}g_{\alpha\gamma}(-1)^i\epsilon_{\alpha\beta\gamma}S^{\alpha}_iS^{\beta}_jS^{\gamma}_k$ with a coefficient $g_{\alpha\gamma}=O(EB/J_{ij})$~\cite{PRB.103.134444}. The three-spin interaction could also arise from the tunnel current in the STM experiment~\cite{PRX.14.041026}. A staggered magnetic field $h_{\mathrm{st}}$ can lead to a similar perturbation term
$\mathcal{H}_{\mathrm{pert}}\sim\sum_{i,j,k} g(-1)^{i} S^x_i S^y_j S^z_k$ with $g=O(h_{\mathrm{st}}^3/(J_{ij} J_{jk}))$~\cite{PRB.99.224409}. The staggered magnetic field may be regarded as a proximity effect of the weak antiferromagnetic order in some Kitaev candidate materials such as Cu$_2$IrO$_3$ and A$_3$LiIr$_2$O$_6$ (A $=$ Cu, Ag)~\cite{PRL.102.017205, PRB.100.094418, doi:10.1021/jacs.7b06911, PRX.9.031047,doi.org/10.1002/pssb.202100146, PRL.108.127203, PRL.123.237203, C9DT01789E},
and also in a hetrostructure of a Kitaev material and an antiferromagnet~\cite{PRB.105.165152}. Although details of these perturbations are different, they can be reduced to a same expression based on the Majorana fermion description. To be concrete, we focus on the staggered Zeeman term, although the other perturbations can be treated in a similar manner.
By representing the spin operators in terms of Majorana fermions, $S^{\alpha}_i = i b^{\alpha}_i c_i$, the Hamiltonian $\mathcal{H}=\mathcal{H}_K+\mathcal{H}_{\mathrm{pert}}$ can be written within the perturbation theory as
\begin{align}
    \mathcal{H}_K &= i \sum_{\ev{i,j}} J_{ij} u_{ij} \, c_i c_j , \\
    \mathcal{H}_{\mathrm{pert}} &= i \sum_{i,j,k} d_{ik} \frac{h_{\mathrm{st}}^3}{J_{ij} J_{jk}} \, u_{ij} u_{jk} \, c_i c_k ,
    \label{eq::majorana_Hamiltonian}
\end{align}
where $c_i$ denotes the itinerant Majorana fermion and $b_i^\alpha$ the gauge Majorana. The $\mathbb{Z}_2$ flux is given by $u_{ij} = i b_i^\alpha b_j^\alpha = \pm 1$. We have neglected $O(1)$ perfactors in Eq.~\eqref{eq::majorana_Hamiltonian} for simplicity.  The coefficient $d_{ik} = \pm 1$ is determined by the orientation of the next-nearest-neighbor hopping shown in Fig.~\ref{fig::majorana_scketch}(b); it takes $+1$ when the hopping direction coincides with the arrow and $-1$ otherwise.
In the absence of the staggered Zeeman magnetic field, $h_{\mathrm{st}}=0$, and with the lattice constant set to $a_0=1$, the band structure of the Majorana fermions hosts the Dirac points at $K = \left(4\pi/(3\sqrt{3}),\, 0 \right), \qquad K^{\prime} = \left(2\pi/(3\sqrt{3}),\, 2\pi/3 \right)$. For $h_{\mathrm{st}}>0$, time-reversal and inversion symmetries are broken: the Dirac point at the $K$ point is shifted to positive energies, while that at the $K'$ point is to negative energies. As a result, small Fermi surfaces of Majorana quasiparticles are formed in the vicinity of each Dirac point (Fig.~\ref{fig:BZ}).
It is known that the ground state of the Kitaev spin liquid belongs to the flux-free sector~\cite{PRL.73.2158}.
Therefore, we basically perform numerical calculations in the flux-free sector by setting $u_{ij}=+1$ on every bond. When necessary, we also calculate the two-flux sector, which we obtain by flipping the sign of a single bond variable, $u_{kl}=-1$, while keeping $u_{ij}=+1$ on all the other bonds.

\noindent\magenta{\it Kitaev spin liquid with triaxial strain.}
We introduce lattice strain into this model to generate pseudo-Landau levels.
It is known that, when an appropriate strain pattern is applied, the particles near the Dirac points couple to an effective vector potential $\bm{A}_{\mathrm{eff}}$ and cause Landau quantization~\cite{VOZMEDIANO2010109, PRB.65.235412, PRB.93.035456, Guinea2010}. In particular, for a kagome lattice with Fermi surfaces located in the vicinity of the Dirac points, strain-induced Landau quantization gives rise to quantum oscillations~\cite{PRB.102.045151}. On the honeycomb lattice, a Landau quantization can be realized by applying triaxial strain. The corresponding displacement vector is given by
\begin{align}
    \bm{U}(x, y) = C(2xy,\, x^2 - y^2) ,
\end{align}
where $x$ and $y$ are the coordinates of real-space , and $C$ is the strength of the strain. The strain tensor is defined as $ \epsilon_{ij} = \frac{1}{2} \left( \partial_i U_j + \partial_j U_i \right)$. It is known that the quasiparticles near the Dirac points couple to the effective vector potential $\bm{A}_{\mathrm{eff}} \propto ( \epsilon_{xx}-\epsilon_{yy}, -2 \epsilon_{xy})= C(4y, -4x)$ generated by the strain. 
The corresponding persuade-magnetic field $B_{\mathrm{eff}} = (\nabla \times \bm{A}_{\mathrm{eff}})_z \propto C$ acts on the itinerant Majorana fermions in the present system.

When finite strain is applied to the honeycomb lattice, the Kitaev interaction is modulated as
\begin{align}
    J_{ij} = J \left( 1 - b \left(|\bm{\delta}_{ij}|/a_0 - 1 \right) \right),
\end{align}
where $\bm{\delta}_{ij} = \bm{R}_i + \bm{U}_i - \bm{R}_j - \bm{U}_j$ is the distorted bond vector, $b$ is the magnetoelastic coupling constant, and $a_0=1$ is the lattice constant. For simplicity, we set $b=1$ throughout this work. $J=1$ is the energy unit. We note that the effective vector potential $\bm{A}_{\text{ eff}}$ induced by the triaxial strain is distinct from Zeeman magnetic fields. The effective vector potential $\bm{A}_{\text{ eff}}$  is coupled to the orbital motion of the emergent itinerant Majorana fermions.

In our numerical calculations, we consider a hexagonal flake of radius $r$, obtained by concentrically extending the honeycomb lattice by $r$ layers from the center, as illustrated in Fig.~\ref{fig::majorana_scketch}(a), and impose open boundary conditions. Let $N$ be the number of unit cells in the honeycomb lattice. For a flake of radius $r$, the total number of sites is related to $N$ as $2N = 6r^2$. The maximal strain $C_{\mathrm{max}}$ is estimated from the value at which the Kitaev coupling $J_{ij}$ at the edge of the flake starts to change its sign. Previous studies have shown that such a sign change occurs for $C\, r \gtrsim 0.3$~\cite{PRL.116.167201}; therefore, we restrict ourselves to the regime $C < 0.3/r$. 

\begin{figure}[t]
  \centering
  \includegraphics[width=1.0\columnwidth]{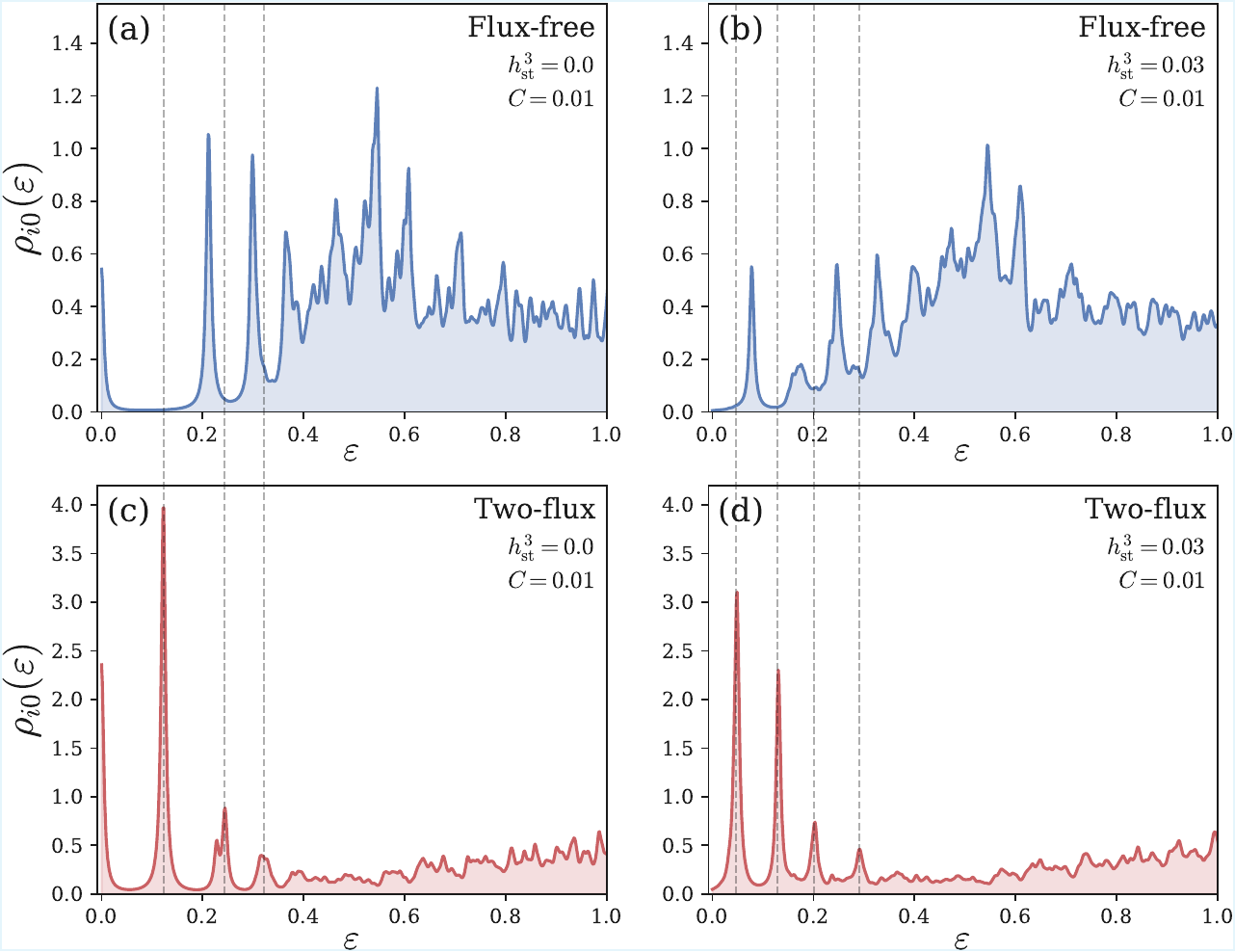}
  \caption{
  Local density of states (LDOS) $\rho_{i_0}(\varepsilon)$ at the site $i_0$ located near the center of the flake.
  The flake radius is $r=20$ and the strain strength is $C=0.01$. We have approximated the delta function by a Lorentzian with broadening width 0.005. (a) and (b): results in the flux-free sector for $h_{\mathrm{st}}^3 = 0.0$ and $h_{\mathrm{st}}^3 = 0.03$, respectively. (c) and (d): results in the two-flux sector for $h_{\mathrm{st}}^3=0.0$ and $h_{\mathrm{st}}^3$, respectively. The black dashed lines indicate that the peaks in the two-flux sector are located between the corresponding peaks in the flux-free sector.
  }
  \label{fig::LDOS}
\end{figure}

\noindent\magenta{\it Pseudo-Landau levels in Majorana-fermi surface state.}
The formation of pseudo-Landau levels can be confirmed by calculating the local density of states (LDOS) and analyzing the single-particle spectrum.
The LDOS at the site $i$ is defined as
\begin{align}
    \rho_i(\varepsilon) = \sum_{n} |\Psi_{n i}|^2 \delta(\varepsilon - \varepsilon_n) ,
\end{align}
where $\varepsilon_n$ is the $n$-th eigenenergy and $|\Psi_{n i}|^2$ is the amplitude of the corresponding eigenstate at the site $i$. In this work, we consider a hexagonal cluster with flake radius $r=20$, and compute the LDOS at a site $i_0$ located near the center of the flake in the flux-free sector. Other system sizes are discussed in Supplemental Materials~\cite{SM}. The results are shown in Figs.~\ref{fig::LDOS}(a) and (c). Focusing on a low-energy region, one finds that the energy levels appear as discrete peaks, indicating the formation of pseudo-Landau levels~\cite{PRL.116.167201}. A particularly notable feature is that the Landau zero-mode ($\varepsilon_n\simeq0$) present at $h_{\mathrm{st}}^3 = 0$ survives even in a staggered Zeeman magnetic field $h_{\mathrm{st}}>0$, although the peak position is shifted from the exact zero energy. This suggests that, in the Majorana-Fermi surface state, the low-energy excitations near the Fermi surface retain their Dirac-like character.
It is also noted that, for $h_{\mathrm{st}} >0$, height of the peak corresponding to each Landau level is reduced.
This behavior can be understood from the band structure at zero strain ($C=0$). When the staggered Zeeman magnetic field $h_{\mathrm{st}}$ is applied, the Dirac points at $K$ and $K^{\prime}$ are shifted upward and downward in energy by $ \Delta \varepsilon \propto \pm h_{\mathrm{st}}^3$, This energy shift manifests itself as a reduction in the peak height of the LDOS.

Furthermore, the analysis of the LDOS in the two-flux sector provides additional evidence for Landau quantization in the Majorana-Fermi surface state. Specifically, we introduce a two-flux configuration by setting $u_{ij} = -1$ on a single bond adjacent to the site $i_0$, while keeping $u_{ij}=+1$ on all other bonds, and calculate the LDOS in this sector. In the Kitaev spin liquid at $h_{\mathrm{st}} = 0$, it is known that, in the two-flux sector, localized excitation modes that do not appear in the flux-free sector emerge between successive Landau levels~\cite{PRL.116.167201}. As shown in Figs.~\ref{fig::LDOS}(b) and (d), such nontrivial bound states appearing between the Landau levels are also present in the Majorana-Fermi surface state at $h_{\mathrm{st}}>0$. This phenomenon indicates that the characteristic features of Landau quantization in the zero-field Kitaev spin liquid are preserved even in the Majorana-Fermi surface state.

\begin{figure}[tb]
  \centering
  \includegraphics[width=1.0\columnwidth]{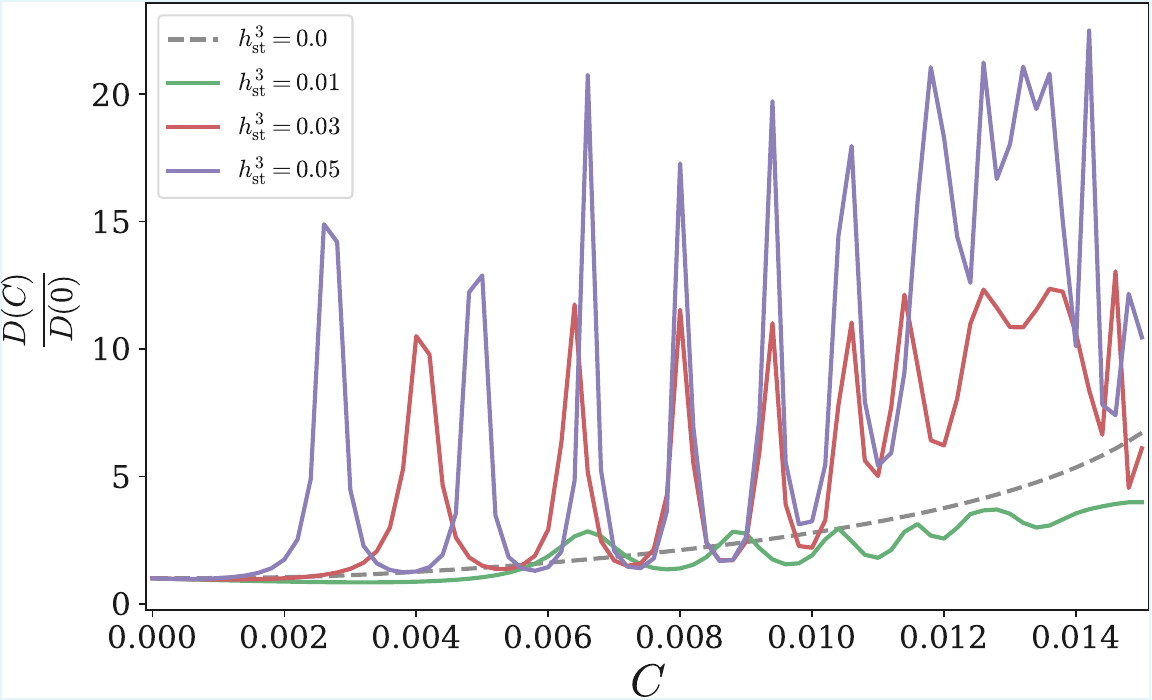}
  \caption{Strain dependence of the normalized DOS $D(C)/D(C=0)$ at zero energy $\varepsilon=0$. The radius is $r=20$. We have approximated the delta function by a Lorentzian with broadening width 0.005. The black dashed line corresponds to $(h_{\mathrm{st}})^3 = 0.0$, while the green, red and purplelines are for $(h_{\mathrm{st}})^3 = 0.01$, $0.03$, and $0.05$, respectively.
  }
  \label{fig::DOS}
\end{figure}

\noindent\magenta{\it Quantum oscillations of the DOS and specific heat.}
Quantum oscillations associated with the Landau quantization of itinerant Majorana fermions appear prominently in the total density of states (DOS) $D(\varepsilon)=\sum_n \delta(\varepsilon - \varepsilon_n)$. Figure~\ref{fig::DOS} shows the strain dependence of the DOS at zero energy $\varepsilon=0$. For $h_{\mathrm{st}} = 0$, the DOS at $\varepsilon=0$ increases almost monotonically with increasing the strain, since a Landau zero-mode is formed almost at the Dirac point. In contrast, for $h_{\mathrm{st}} > 0$, clear oscillatory structures emerge in the DOS. The peaks in the DOS correspond to successive level crossings of the energy spectrum at $\varepsilon=0$~\cite{SM}. More precisely, the DOS $D(\varepsilon=0)$ is enhanced when an eigenenergy flows e.g. from $\varepsilon<0$ to $\varepsilon>0$ as a function of the strain $C$, which is exactly what happens in metals with Fermi surfaces. The oscillation amplitude increases as $h_{\mathrm{st}}$ becomes large, since the size of the Majorana-Fermi surface grows as $h_{\mathrm{st}}$ increases.
Therefore, the oscillations in the present system can be interpreted as quantum oscillations arising from the Landau quantization of the Majorana-Fermi surface, i.e., ``Majorana quantum oscillations.'' 

The Majorana quantum oscillations in the Majorana-Fermi surface state can also be seen from the behavior of the specific heat. For $h_{\mathrm{st}}=0$, it is known that the ground state of the Kitaev spin liquid belongs to the flux-free sector, whereas at finite temperatures $\mathbb{Z}_2$ flux excitations are thermally excited. On the other hand, at sufficiently low temperatures, the flux density is strongly suppressed and the system remains close to the flux-free sector~\cite{PRL.113.197205}. Therefore, the low-temperature thermodynamic properties can be approximated by taking into account only the contribution from the Majorana fermions. In this case, the specific heat $\mathcal{C}_v$ can be expressed in the flux-free sector as
\begin{align}
    \mathcal{C}_v = \frac{\beta^2}{2N} \sum_{n} \varepsilon_n^2 f(\varepsilon_n)\bigl(1 - f(\varepsilon_n)\bigr) ,
\end{align}
where $\beta = 1/T$ is the inverse temperature and $f(\varepsilon_n) = 1/\bigl(1 + e^{\beta \varepsilon_n}\bigr)$ is the Fermi distribution function.

\begin{figure}[tb]
  \centering
  
  \includegraphics[width=1.0\columnwidth]{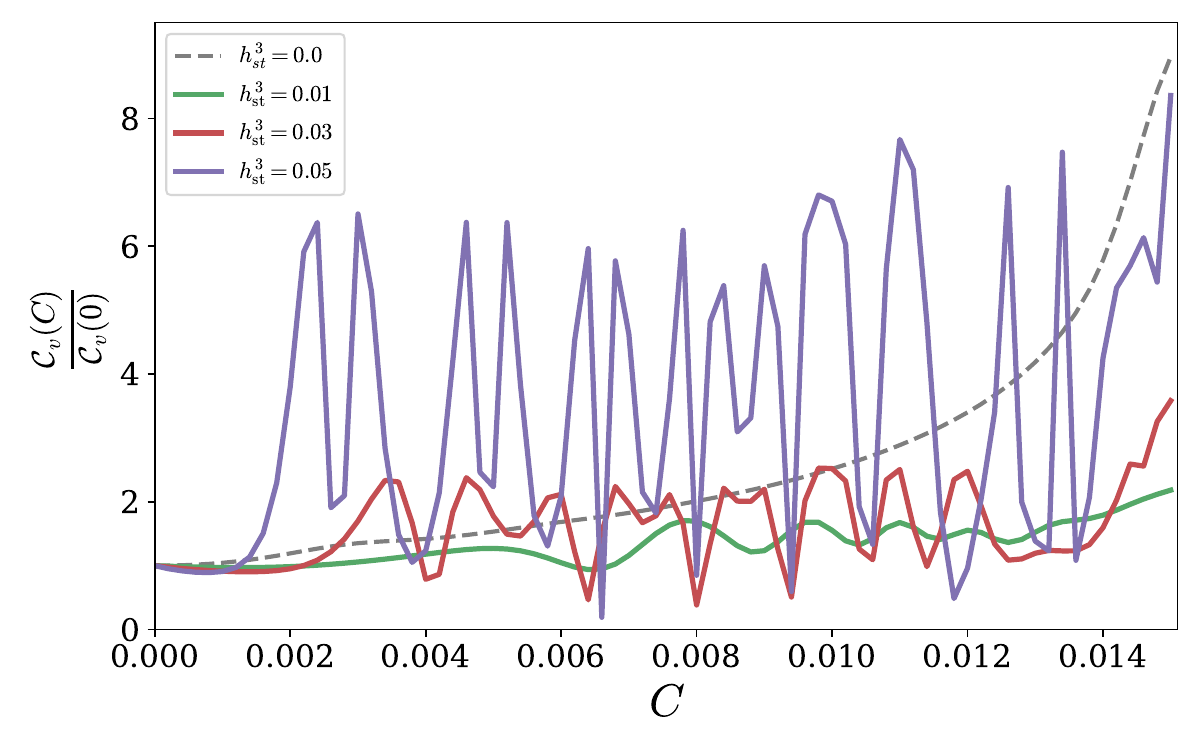}
  \caption{Normalized specific heat $\mathcal{C}_v(C)/\mathcal{C}_v(C=0)$ as a function of strain for a hexagonal flake with radius $r=20$.
The black dashed line corresponds to $h_{\mathrm{st}}^3 = 0.0$, while the green, red and purple lines are for $h_{\mathrm{st}}^3 = 0.01$, $0.03$, and $0.05$, respectively.
}
\label{fig::Cv}
\end{figure}

We fix the temperature $T = 0.001$ and investigate the specific heat under strain (Fig.~\ref{fig::Cv}). Similarly to the DOS, at $h_{\mathrm{st}}=0$, the specific heat monotonically increases as the strain is applied. In sharp contrast, for $h_{\mathrm{st}} > 0$, clear oscillations are induced by the strain and the oscillation amplitude gets larger when $h_{\mathrm{st}}$ increases. This behavior is analogous to the de Haas-van Alphen effect in metals with Fermi surfaces and can be interpreted as Majorana quantum oscillations originating from the Landau quantization of the Majorana-Fermi surface.

\noindent\magenta{\it Discussion on experiments.}
We finally discuss possible experimental observations of the Majorana quantum oscillation.
One can realize a Majorana-Fermi surfaces in a Kitaev system with a simple staggered antiferromagnetic order or stacked zigzag antiferromagnetic layers on top of the spin liquid state~\cite{PRB.99.224409,PRB.105.165152}. Majorana-Fermi surface states can also be generated in the presence of external perturbations such as uniform electromagnetic fields and local STM currents~\cite{PRB.103.134444, PRX.14.041026, PRB.78.024402}. 
For the STM setup, since the Majorana-Fermi surfaces only emerge in the region where the current flows, bulk thermodynamic quantities would not show quantum oscillation. However, this setup has the advantage that one can directly probe local quantities such as the LDOS, which should display clear signatures of Majorana quantum oscillations.

\noindent\magenta{\it Conclusions.}
In this work, we have investigated the Landau quantization and quantum oscillations in the Majorana-Fermi surface state. To this end, we considered the $S=1/2$ honeycomb Kitaev model with the perturbation which generates Majorana-Fermi surfaces in the vicinity of the Dirac points, and further applied triaxial strain which induces an effective vector potential acting on the itinerant Majorana fermions. We have shown that the resulting pseudo-Landau levels resemble Dirac fermion systems with finite doping.
From the analysis of the LDOS, we found that a clear sequence of Landau levels appears at sites near the center of the flake, and that in the two-flux sector nontrivial modes specific to the Kitaev spin liquid emerge between the Landau levels in the flux free sector.
This suggests that Landau quantization induced by lattice strain can be used to detect the presence of a charge-neutral Majorana-Fermi surface through local and thermodynamic measurements.

\noindent{\it Note added.}
We became aware of a related study on strain-induced Landau quantization in a Majorana-Fermi surface state during the preparation of this manuscript~\cite{zhu2025_arxive}. Their model describes a Majorana-Fermi surface in a uniform magnetic field and does not host any Dirac cones. Hence, their work addresses a different physical situation.

\noindent{\it Acknowledgment.}
This work was supported by JST SPRING Grant No. JPMJSP2132 and JSPS KAKENHI Grant No. 22K03513

\bibliography{ref}%

\clearpage
\onecolumngrid
\section{Supplemental Materials}
\twocolumngrid

\subsection{A. Energy spectrum of Majorana Hamiltonian}
\begin{figure}[htb]
  \centering
  \includegraphics[width=\linewidth]{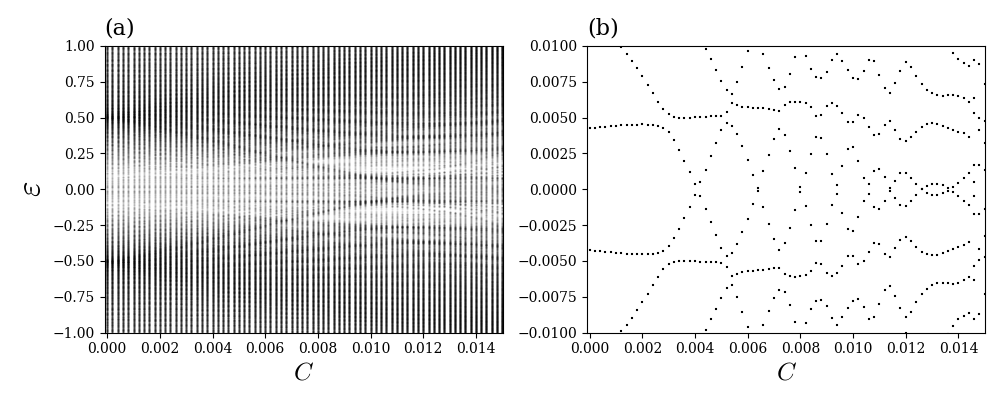}
  \caption{Strain dependence of the energy spectrum for (a) a wide energy region and (b) a low-energy part with $|\varepsilon| < 0.01$.
  The calculations are performed for $h_{\mathrm{st}}^3 = 0.03$ and $r=20$.
}
\label{fig::enegy_spectrum_r=20_h003}
\end{figure}

The nature of the Majorana quantum oscillations can also be understood from the behavior of the energy spectrum. 
For $h_{\mathrm{st}} > 0$, we find Landau-quantization–induced level crossings among the low-energy levels in the vicinity of $\varepsilon = 0$ (Fig.~\ref{fig::enegy_spectrum_r=20_h003}). This is precisely the origin of the Majorana quantum oscillations. When an energy eigenvalue flows crossing the zero energy, the DOS at $\varepsilon=0$ increases and consequently exhibits oscillations. In particular, the crossings occur around $C \sim 0.004$, $0.006$, and $0.008$, which approximately coincide with the peak positions of the quantum oscillations in the DOS and in the specific heat (Figs.~\ref{fig::DOS} and \ref{fig::Cv} in the main text). We note that the peak in the specific heat splits into two, suggesting a possible connection to an additional avoided level crossing at $\varepsilon \sim 0.005$.
Note also that spectral flows with level crossings do not take place for the pure Kitaev spin liquid state at $h_{\mathrm{st}}=0$ without a Fermi surface. 


\subsection{B. System size dependence of Majorana quantum oscillations}
\begin{figure}[htb]
  \centering
  \includegraphics[width=1.0\columnwidth]{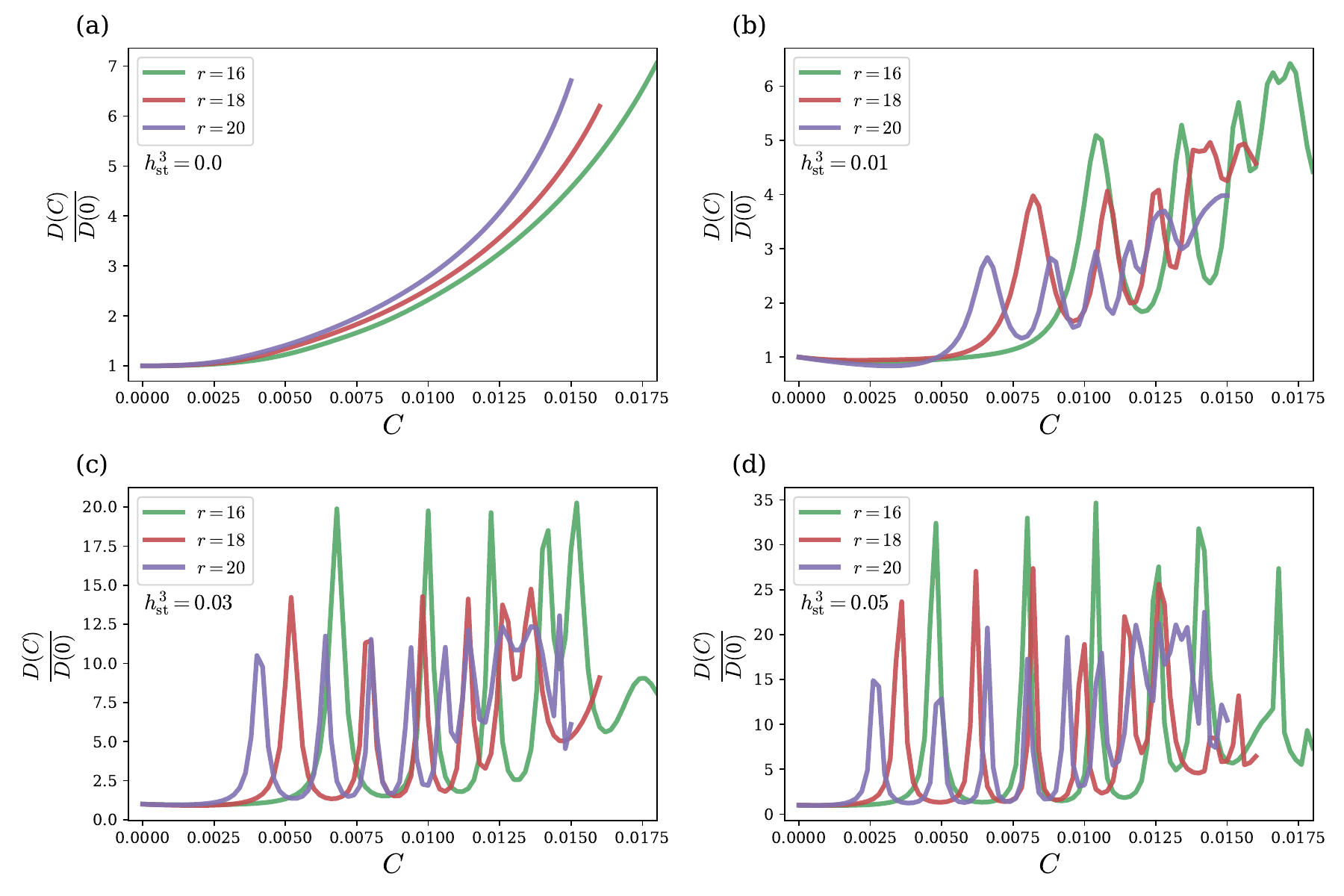}
  \caption{Strain dependence of the normalized DOS $D(C)/D(C=0)$ at zero energy $\varepsilon=0$.
  The calculations are performed for $r=16$ (green), $r=18$ (red) and $r=20$ (purple).
  (a) $h_{\mathrm{st}}^3=0$, (b) $h_{\mathrm{st}}^3=0.01$, (c) $h_{\mathrm{st}}^3=0.03$, and (d) $h_{\mathrm{st}}^3=0.05$. 
}
\label{fig::DOS_r}
\end{figure}

\begin{figure}[htb]
  \centering
  \includegraphics[width=1.0\columnwidth]{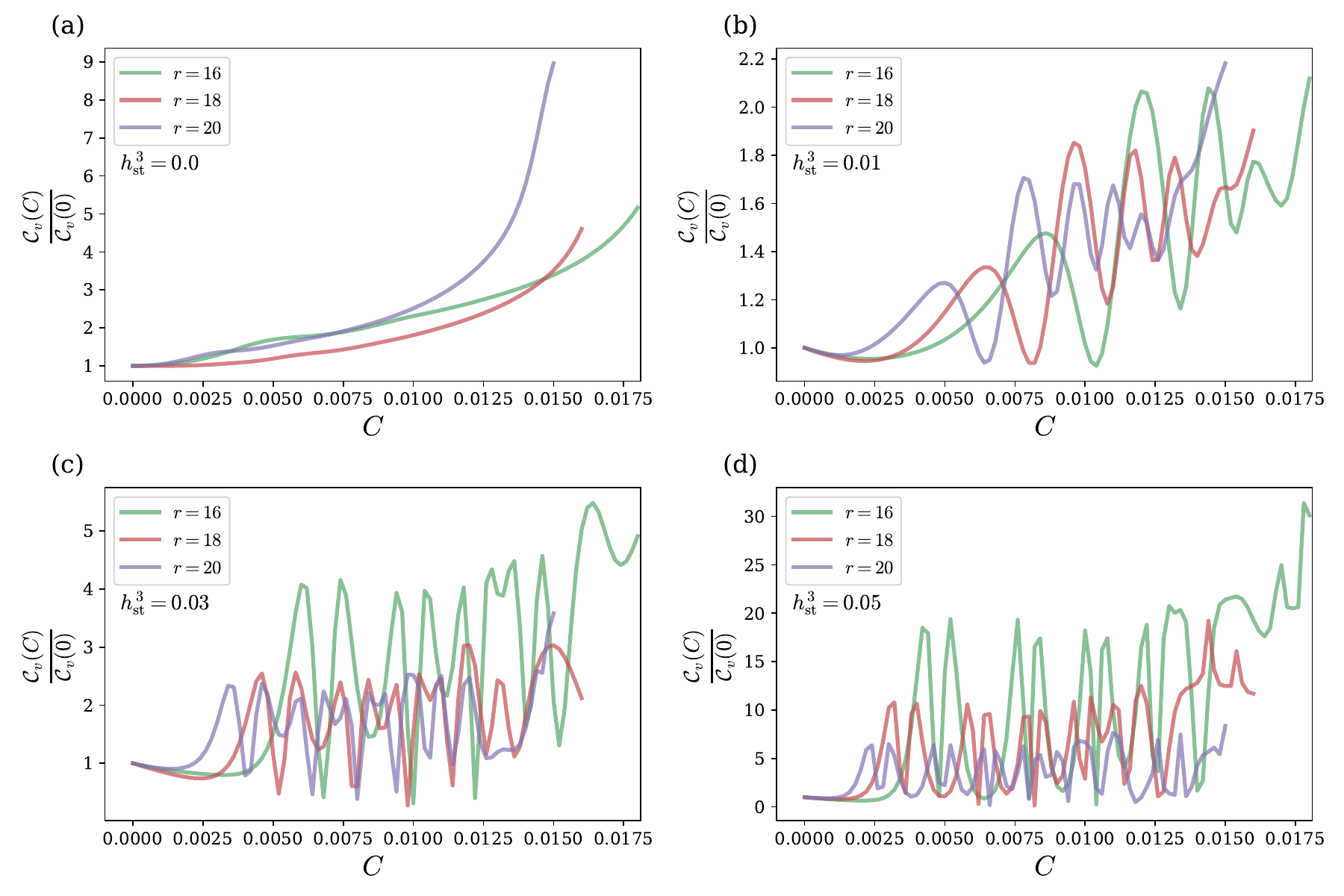}
  \caption{Strain dependence of the normalized specific heat $\mathcal{C}_v(C)/\mathcal{C}_v(C=0)$ at $T=0.001$.
  The calculations are performed for $r=16$ (green), $r=18$ (red) and $r=20$ (purple).
  (a) $h_{\mathrm{st}}^3=0.0$, (b) $h_{\mathrm{st}}^3=0.01$, (c) $h_{\mathrm{st}}^3=0.03$, and (d) $h_{\mathrm{st}}^3=0.05$. 
}
\label{fig::Specific_heat_r}
\end{figure}

We perform calculations for several system sizes to confirm that our results for the Majorana quantum oscillations are robust. We show the DOS $D(C)$ at $\varepsilon=0$ and specific heat $\mathcal{C}_v(C)$ at $T=0.001$ for the system sizes $r = 16, 18, 20$ (Figs.~\ref{fig::DOS_r} and \ref{fig::Specific_heat_r}). For $h_{\mathrm{st}}=0$ where there is no Majorana-Fermi surface, $D(C)$ and $\mathcal{C}_v(C)$ monotonically increases and the Majorana quantum oscillations are absent for all the system sizes. On the other hand, the Majorana quantum oscillations clearly appear for $h_{\mathrm{st}} > 0$ independently of the system size. The oscillation amplitude is suppressed for large system sizes. This is because the ratio of the number of the level crossing eigenenergies to that of the low eigenenergies within the energy window $|\varepsilon_n|<T$ is small for large system sizes, and consequently effects of the spectral flow of each eigenenergy become weak. This is common also to the conventional metals.
Although it may be favorable to use a small flake in an experiment to detect the Majorana quantum oscillation for this reason, we expect it is measurable in  relatively large samples since the de Haas-van Alphen effect which has similar finite size effects is observed usually in bulk systems.


\end{document}